\documentclass[a4paper]{article}
\usepackage{epsfig,amsmath,amsfonts,amssymb,setspace,textcomp}
\usepackage[T1]{fontenc}
\newcommand{\ms}{ms$^{-1}$}

\newcommand{\me}{M$_{\rm{\oplus}}$}
\makeatletter
\renewcommand{\@oddhead}{\textit{Advances in Astronomy and Space Physics} \hfil}
\renewcommand{\@evenfoot}{\hfil \thepage \hfil}
\renewcommand{\@oddfoot}{\hfil \thepage \hfil}
\makeatother

\renewenvironment{thebibliography}[1]{\begin{oldthebibliography}{#1}\setlength{\parskip}{0ex}\setlength{\itemsep}{0ex}}{\end{oldthebibliography}}
\begin{document}
\def\Tef{$T\rm_{eff }$}
\def\vt{$V_t$}
\def\logg{$\log g$}
\def\HD{HD 77338~}

\fontsize{11}{11}\selectfont % the font size cannot be changed in any case!
%  insert your title, authors information and text instead of the one provided below
\title{Abundances in the atmosphere of the metal-rich planet-host star \\ HD 77338}
\author{\textsl{I.O.Kushniruk$^{1}$, Ya.V.Pavlenko$^{2,3}$, J.S.Jenkins$^{4,2}$, H.R.A.Jones$^2$}}
\date{\vspace*{-6ex}}
\maketitle
\begin{center} {\small $^1$Taras Shevchenko National University of Kyiv, Glushkova ave., 4, 03127, Kyiv, Ukraine\\
$^2$Main Astronomical Observatory of the NAS of Ukraine, 27 Akademika Zabolotnoho St., 03680 Kyiv, Ukraine\\
$^3$Centre for Astrophysics Research, University of Hertfordshire,
College Lane, Hatfield, Hertfordshire AL10 9AB, UK \\
$^4$Departamento de Astronom\'ia, Universidad de Chile,
Camino el Observatorio 1515, Las Condes, Santiago, Chile \\
{\tt nondanone@gmail.com}}
\end{center}

\begin{abstract} Abundances of
Fe, Si, Ni, Ti, Na, Mg, Cu, Zn, Mn, Cr and Ca
in the atmosphere of the K-dwarf HD 77338 are determined and discussed. 
HD 77338 hosts a hot Uranus-like planet and is currently the most metal-rich 
single star to host any planet. 
Determination of abundances 
was carried out in the framework of a self-consistent approach 
developed by Pavlenko et al. (2012).
Abundances were computed iteratively by the program ABEL8, and the process 
converged after 4 iterations.  We find that most elements follow the iron 
abundance, however some of the iron peak elements are found to be over-abundant in 
this star.
\end{abstract}

\section{Introduction}
\indent\indent\indent  Determining the chemical composition of stars is one of 
the primary goals of astrophysics. 
Such investigations help us to better understand the chemical enrichment of the 
Galaxy and to
make some assumptions about the mechanisms involved in element evolution in the 
interstellar medium, and in stellar atmospheres in particular \cite{cu}.
While studying the Sun, the problem of the abundances of certain atoms necessitated  
a model to explain this. 
It was finally 
explained with the introduction of the pp- and CNO- cycles in the interior of the Sun.
But this was not enough to explain the presence of large amounts of helium. 
The next step in studying the evolution of elements was the introduction of nucleosynthesis 
theory. Modern scientific understanding is that chemical 
elements were formed as a result
of the processes occurring in stars, leading to evolutionary 
changes of their physical conditions. 
Therefore, the problem of nuclide formation is also closely related 
to the issue of the evolution of stars and planetary system.
Recently Jenkins et al.
\cite{uranus} announced the discovery of a low-mass planet orbiting the super 
 HD77338 as part of our ongoing Calan-Hertfordshire 
Extrasolar Planet Search \cite{jenkins2013a}. The best-fit planet solution has an orbital 
period of 5.7361 $\pm$ 0.0015 days and with a radial velocity semi-amplitude 
of only 5.96$\pm$1.74~\ms, giving a minimum mass of $^{+4.7}_{-5.3}$~\me. 
The best-fit eccentricity from this solution is 0.09$^{+0.25}_{-0.09}$, 
and is in the agreement with results 
of a Bayesian analysis and a periodogram analysis.

\indent\indent According to modern theory, the formation of the nucleus of chemical elements from carbon to iron 
is the result of thermonuclear reactions involving He, C, O, Ne and Si in stars.
After the depletion of hydrogen reserves, a star's core starts running 
a 3$\alpha $ reaction, where 
it produces a number of elements as a result of the following transformations:
\indent $\mathrm{3\,{}^4He\,\longrightarrow\,{}^{12}C}$,
 $\mathrm{{}^{12}C \,+\, {}^4He \,\longrightarrow\, {}^{16}O \,+\, \gamma}$,
$\mathrm{{}^{16}O \,+\, {}^4He \,\longrightarrow\, {}^{20}Ne \,+\, \gamma}$.\\
 \indent\indent After reaching a specific threshold temperature, carbon begins fusing with the formation of Ne, Na and Mg:
 
\indent $\mathrm{{}^{12}C \,+\, {}^{12}C \,\longrightarrow\, {}^{20}Ne \,+\, {}^{4}He \,+\, 4,62 MeV }$,\\
\indent $\mathrm{{}^{12}C \,+\, {}^{12}C \,\longrightarrow\, {}^{23}Na \,+\, p \,+\, 2,24 MeV }$,\\
\indent $\mathrm{{}^{12}C \,+\, {}^{12}C \,\longrightarrow\, {}^{24}Mg \,+\,\gamma \,-\, 2,60 MeV }$.\\
\indent\indent Aluminum can then be produced by: $\mathrm{{}^{24}Mg \,+\, p \,\longrightarrow\, {}^{25}Al \,+\, \gamma}$.\\
\indent\indent The combustion reaction of oxygen is a dual-channel process and causes the presence of Al, S, P, Si and Mg. One of the channels is: \\
\indent $\mathrm{{}^{16}O \,+\, {}^{16}O \,\longrightarrow\, {}^{30}Si \,+\,{}^{1}H \,+\, {}^{1}H \, +\, 0,39 MeV }$,\\
\indent $\mathrm{{}^{16}O \,+\, {}^{16}O \,\longrightarrow\, {}^{24}Mg \,+\, {}^{4}He \,+\,{}^{4}He \,- \, 0,39 MeV }$,\\
\indent $\mathrm{{}^{16}O \,+\, {}^{16}O \,\longrightarrow\, {}^{27}Al \,+\, {}^{4}He \,+\, {}^{1}H \, -\, 1,99 MeV }$,\\
\indent\indent With continuous temperature growth, silicon burning is initiated. This process is described by a number of reactions. As a result we can receive,
for example, Ar, Ni, S, etc. $\mathrm{{}^{56}Ni}$, after two $ \beta $ decays, turns into $\mathrm{{}^{56}Fe}$. It is the final stage of the 
fusion of nuclides in massive stars, which forms the nucleus of the iron group.\\
\indent\indent The production of heavy elements is provided by other mechanisms. They are called s- and r- processes. 
 \begin{itemize}
\item s-process or slow neutron capture: formation of heavier nuclei by lighter nuclei through successive neutron capture.
The original element in the s-process is $\mathrm{{}^{56}Fe}$. 
The reaction chain ends with $\mathrm{{}^{209}Bi}$. It is thought that s-processes occur mostly in stars on the asymptotic giant branch. 
For the s-process to run, an important condition is the ability to produce neutrons. The main neutron source reactions are:
$\mathrm{{}^{13}C \,+\, {}^{4}He \,\longrightarrow\, {}^{16}O \,+\, n }$, 
$\mathrm{{}^{22}Ne \,+\, {}^{4}He \,\longrightarrow\, {}^{25}Mg \,+\, n }$.
Example of s-process reactions are:
$$^{56}\textrm{Fe} + n \longrightarrow {}^{57}\textrm{Fe} + n \longrightarrow {}^{58}\textrm{Fe} + n 
\longrightarrow {}^{59}\textrm{Fe} \stackrel{\beta^-}{\longrightarrow} {}^{59}\textrm{Co} + n \longrightarrow {}^{60}\textrm{Co}
\stackrel{\beta^-}{\longrightarrow} {}^{60}\textrm{Ni} + n {\longrightarrow}~~\textrm{...}$$
 \end{itemize}
 
Elements heavier than H and He are usually called metals in astrophysics. 
Their concentration is significantly less, relative to hydrogen and helium, but  
they are the source of thousands of spectral lines originating from a star's atmosphere. 
The abundance of iron depends on a stars age and on its position in the galaxy \cite{book}. 
Metal-rich stars are also known to be rich in orbiting giant exoplanets. High metallicity 
appears to be a major ingredient in the formation of planets through core accretion \cite{uranus}. \\

HD 77338 is one of  the most metal-rich stars in the sample of \cite{2008} and in the local Solar neighborhood in 
general.  Its  spectral type is given as K0IV in the Hipparcos Main Catalogue \cite{per}.
However, \HD is not a subgiant, as labeled in Hipparcos \cite{uranus}, 
its mass and radius are smaller than the Sun's: M = 0.93 $ \pm $ 0.05 M$_\odot$,
R = 0.88 $ \pm $ 0.04 R$_\odot$. A parallax of 24.54 $ \pm $ 1.06 mas for HD 77338 means the 
star is located at a distance of 40.75 $\pm$ 1.76 pc. 
Its effective temperature and surface gravity were found 
$T_{eff}$ = 5370 $ \pm $ 80 K, log$ g$ = 4.52 $ \pm $ 0.06 \cite{uranus}. 
More stellar parameters for HD 77338 and detailed information about its 
planetary system are in \cite{uranus}.

Using the Simbad database one can find information on the previous assessments of abundances
in the atmosphere of HD 77338 (see Table~\ref{tab1}). In most cases the authors only 
determine the metallicity of the star, i.e. the iron abundance.
 In turn, we recomputed the abundances of many
elements which show significant absorption lines in the observed spectrum of \HD.

\begin{table}
 \centering
 \caption{Simbad's list of previous assessments of abundances
in the atmosphere of HD 77338 }\label{tab1}
 \vspace*{1ex}
 \begin{tabular}{cccccc}
  \hline
  Teff & $log g$ & [Fe/H]  & CompStar  & Reference \\
  \hline
  5300  & 4.30 & 0.36  & Sun &  \cite{prug}\\
  5290  & 4.90 & 0.22  & Sun &  \cite{felt}\\
  5290  & 4.60 & 0.30  & Sun &  \cite{th}\\
 \hline 
 \end{tabular}
\end{table}

\begin{figure}[!h]
\centering
\begin{minipage}[t]{.75\linewidth}
\centering
\epsfig{file = 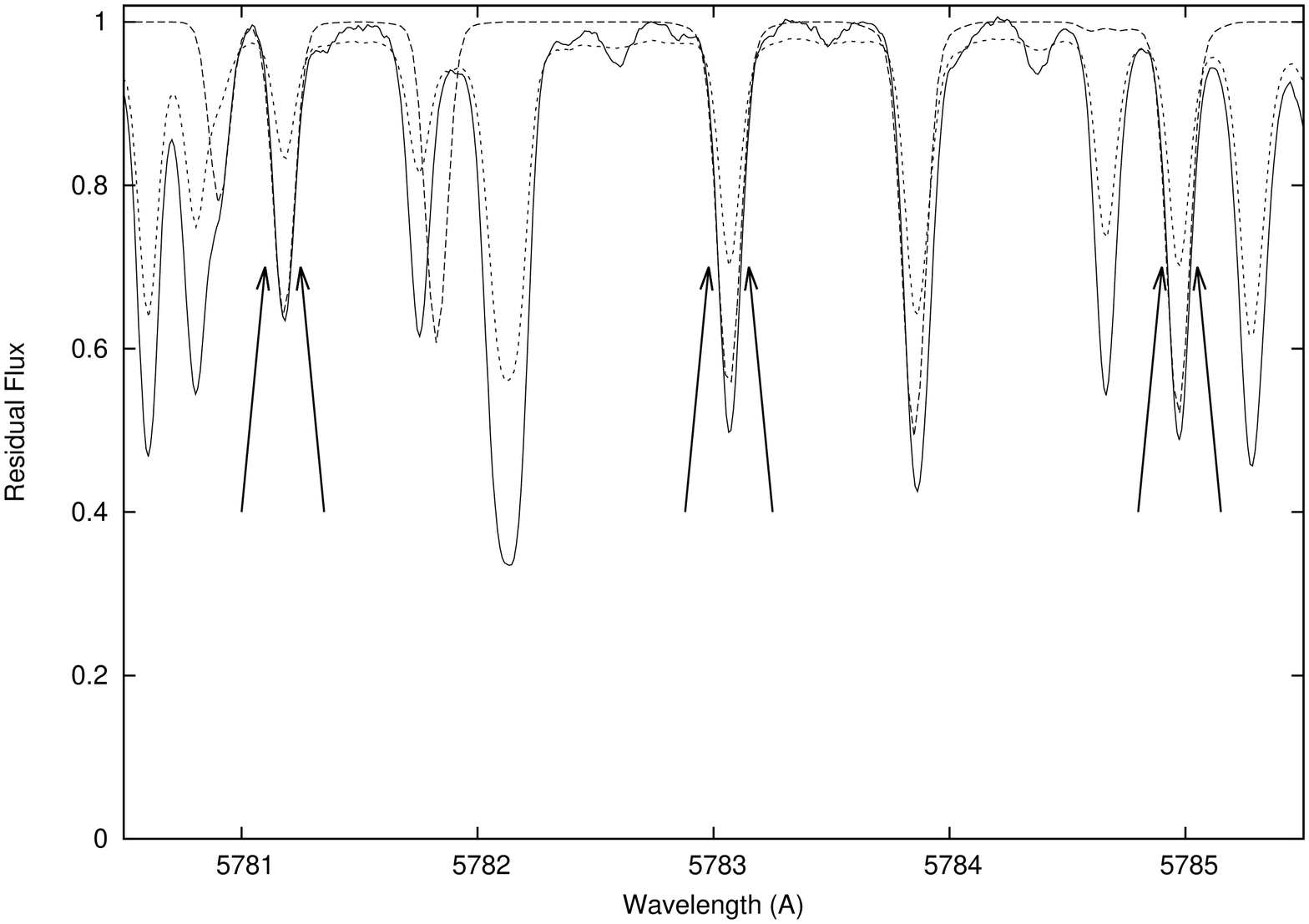,width = .7\linewidth, angle=270}
\caption{The dotted line represents the observed spectrum of the Sun, 
the solid line is the observed spectrum of HD 77338, the dashed
line shows the synthetic spectrum computed by Wita618 with $log$ N(Cr)= $-$ 6.11
for a model atmosphere of 5315/4.39. This plot was used
to detect "clean" parts of Cr line  profiles marked here 
by arrows}\label{fig1}
\end{minipage}
\hfill
\begin{minipage}[t]{.75\linewidth}
\centering
\epsfig{file = 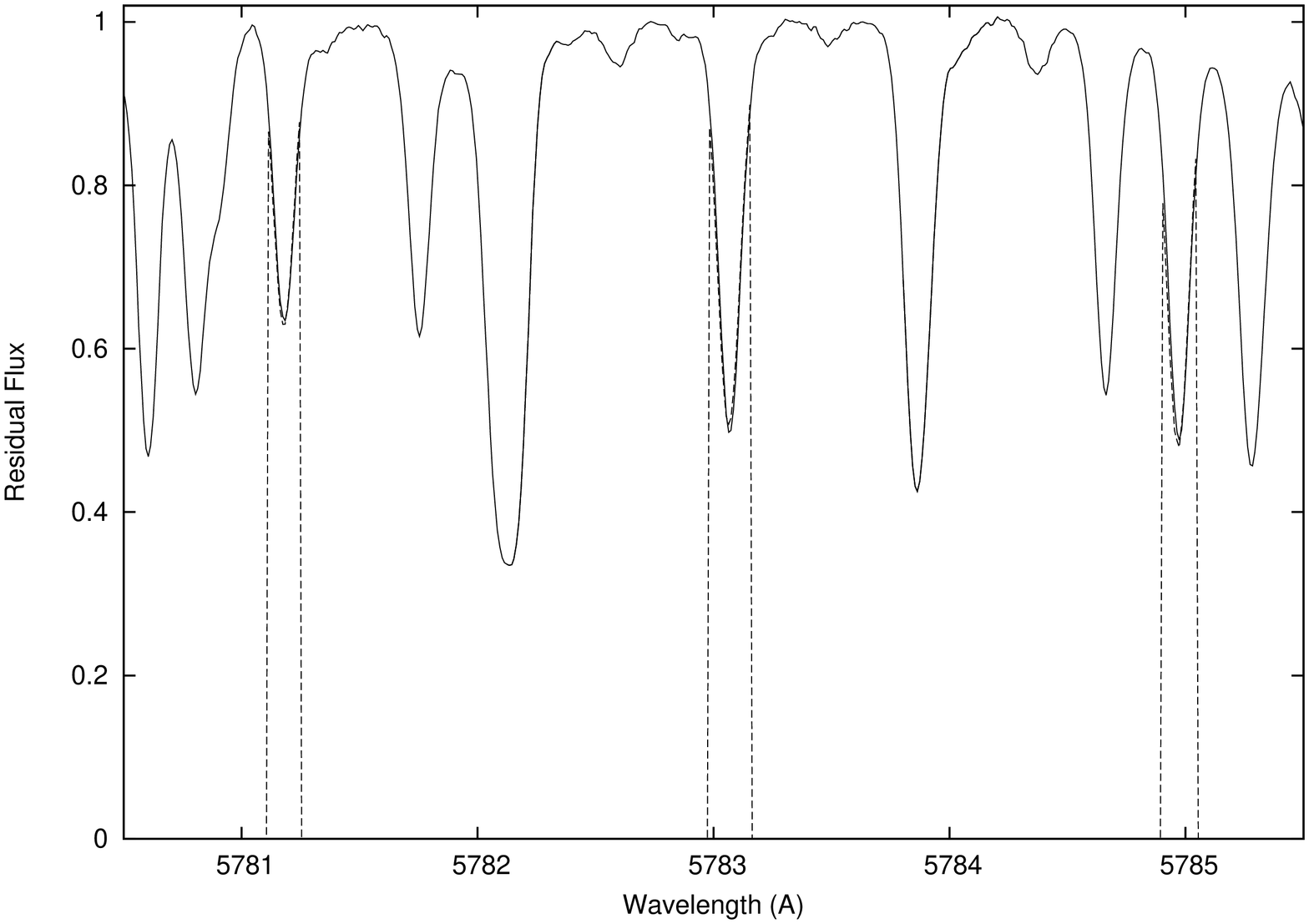,width = .7\linewidth, angle=270}
\caption{Cr I absorption line profiles computed 
for a model atmosphere of 5315/4.39 and $log$ N(Cr) = $-$ 6.09   
using ABEL8 and fitted to 
the observed spectrum of HD 77338 shown by dashed and solid lines,
respectively. The vertical lines show the fitted parts of Cr I line
profiles.}\label{fig2}
\end{minipage}
\end{figure}

\begin{figure}[!h]
\centering
\begin{minipage}[t]{.45\linewidth}
\centering
\epsfig{file = 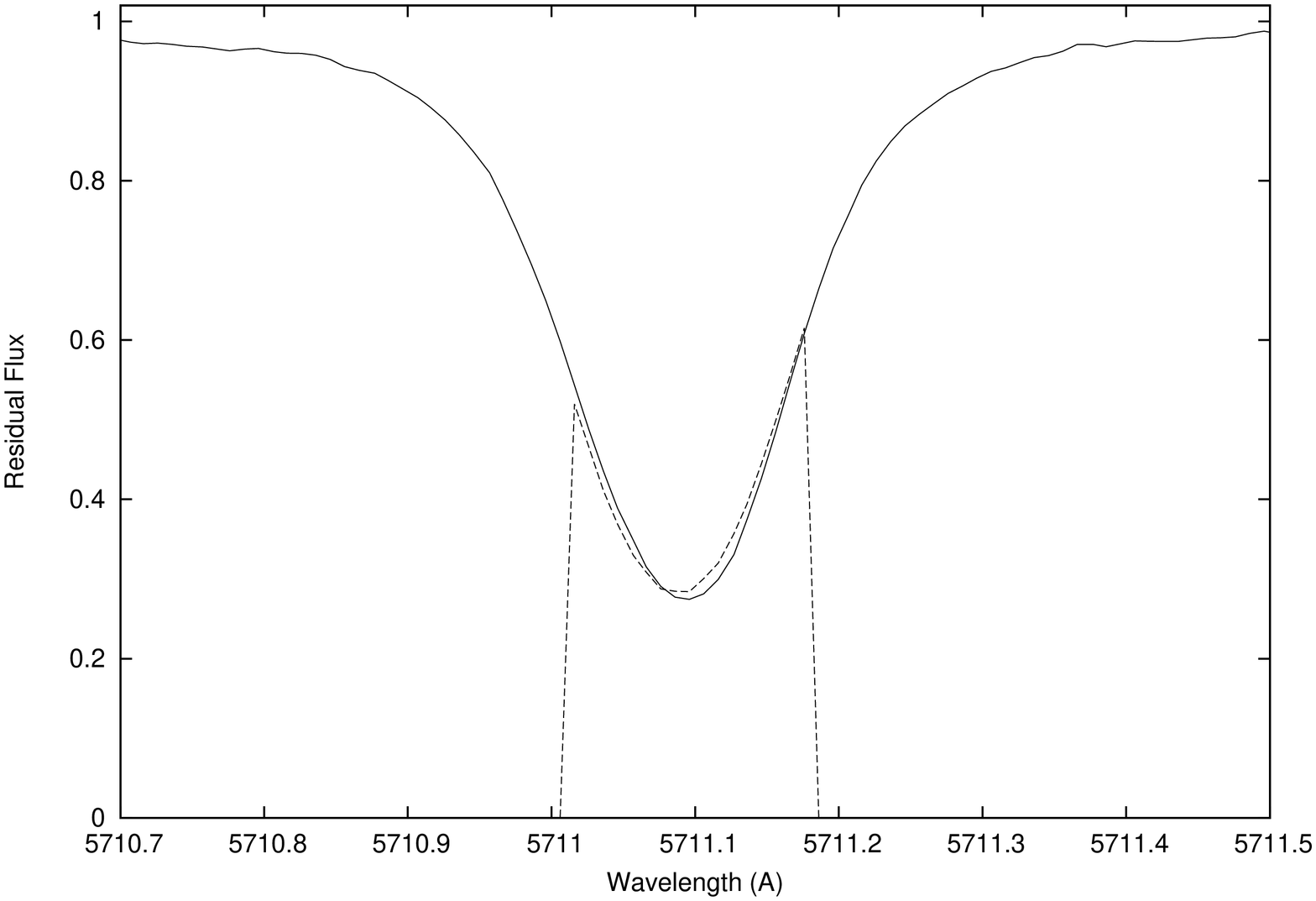,width = .7\linewidth,angle=270}
\caption{Computed by ABEL8 and observed line profiles of Mg in the spectrum of 
HD 77338 with a model atmosphere of 5315/4.39, $log$ N(Mg) = $-$ 4.23
shown by dashed and solid lines, respectively. The vertical lines show 
the fitted part of Mg I line
profile.}\label{fig3}
\end{minipage}
\end{figure}

\section{The observations} 

The observations of \HD were carried out as part of
the Calan-Hertfordshire Extrasolar Planet Search (CHEPS) program \cite{2009}. The  main aim 
of the program  is monitoring a
sample of metal-rich stars in the
southern hemisphere to search for short period planets that have a
high probability to transit their host stars, along with improving the
existing statistics for planets orbiting solar-type and metal-rich stars.
The high-S/N ($ > 50$) and high-resolution ($R = 100\, 000$) spectrum of 
\HD, observed with the HARPS spectrograph \cite{mayor},
was reduced using the standard automated HARPS pipeline and 
analyzed in this work in order to determine the chemical abundances
and other physical parameters of the stellar atmosphere.

\section{The procedure} 

Firstly, we selected ``good'' absorption lines for all elements
of interest that are present in spectra of the Sun and HD 77338.
These lines should be not be blended (see \cite{2008}) and be intense enough in 
both spectra. We selected lists of lines of each element that were to be used 
for the abundance investigations.
We used line list data, which was taken from database of atomic 
absorption spectra VALD \cite{_kupka99}, to compute synthetic spectra of the Sun
for a plane-parallel model atmosphere
with parameters \Tef/\logg/[Fe/H] = 5777/4.44/0.0 \cite{SAM12}.
The model atmosphere was used to compute
the synthetic spectra using WITA6 \cite{97}, building a grid of 
models with different microturbulent velocities \vt = 0 -- 3 km/s with a step size of 0.25 km/s.
The shape of the line absorption profiles were constructed as 
Voigt function profiles $H(a,v)$, and a 
classical approach was used to compute the damping effects \cite{_unsold56}. 
To compute the rotational profile we followed the procedure described in Gray \cite{_gray76}.

All abundance determinations were performed by the ABEL8 program \cite{abel8}.
Details of the full procedure we used is described in \cite{mnras}, see \cite{2008} also 
for more details on the line selection and fitting procedure.

\section{Results}

\subsection{The Sun}

The solar spectrum is well-studied and abundances for the Sun are known 
to very high accuracy, therefore it represents a very good template.
Fig.~\ref{fig1} and Fig.~\ref{fig2} illustrate the presence of
spectral lines of Cr in the observed spectrum of the Sun as a star
\cite{_kurucz84}.
Arrows on the plot show
the spectral range
which was selected to compute profiles of two Cr I lines to be used later by
ABEL8 \cite{abel8} in the determination of the abundance of chromium.
We employ a similar selection in the solar spectrum Sun and the spectrum of HD 77338 for 
lines of Fe I, Si I, Ni I, Ti I, Na I, Mg I, Cu I, Zn I, Mn I, Cr I, Al I, and Ca I.

We verified whether our input data 
are good enough to reproduce the abundances in the atmosphere of the Sun.
We computed abundances for the Sun using the fits of the theoretical spectra to the profiles of the
selected lines. In that way we can test our method and estimate the accuracy of our abundance determination.
Then we investigated the
dependence $E_a = \partial a/\partial E''$, where $a$ and $E''$ are the iron
abundance and excitation potential of the correspondent
radiative transition forming the absorption line.
 Best fits of the selected lines of Fe I in the computed spectra, when compared to 
their observed profiles in the solar spectrum, provides the min  $E_a$ of 
\vt =0.75 km/s. The abundances of iron and other elements were then 
obtained using this adopted value for the microturbulence, the results are shown 
in Table~\ref{tab2}. It is worth noting that our abundances agree with the
reference values within an accuracy of $\pm$0.1 dex.

\subsection{HD 77338}

The model atmosphere for \HD was computed using the parameters 
determined by Jenkins et al.
\cite{uranus} using the SAM12 program \cite{SAM12}.
Again, as the first step of our analysis we determined the 
microturbulent velocity in the atmosphere of \HD.
The minimum of the slope of $E_a$ provides \vt = 0.75 km/s for
$log$ N(Fe) = $-$ 4.120 $ \pm $ 0.07 or [Fe/H] = 0.281 (iteration 1). 
For other elements we used the same value ($V_{t}$ = 0.75 km/s). 

We carried out 4 iterations to determine all abundances.
In each next step the abundances from the previous determination were used to recompute 
the model atmosphere by SAM12 \cite{SAM12} and the synthetic spectra.
Each time we are approaching self-consistency by computing the model 
atmosphere that relates to the final metallicity of the star.

\begin{table}
 \centering
 \caption{Abundances in the atmosphere of HD 77338, iteration 4}\label{tab2}
 \vspace*{1ex}
 \begin{tabular}{cccccccc}
  \hline
  Iron & $log$ N(X) & $log$ N(X)$_\odot$ ABEL8 & $log$ N(X)$_\odot$ & [X/H] & [X/Fe] & $v \sin i$ ($ km/s $) &N$_l$ \\
  \hline
Al I &  $-$ 5.403 $\pm$ 0.000 & $-$ 5.767 $\pm$ 0.033 & $-$ 5.551 & $+$ 0.148 & $-$ 0.095 & 2.33 $\pm$ 0.44 &3\\
Ca I &  $-$ 5.376 $\pm$ 0.029 & $-$ 5.588 $\pm$ 0.023 & $-$ 5.661 & $+$ 0.285 & $+$ 0.042 & 1.25 $\pm$ 0.13 &14\\
Cr I &  $-$ 6.085 $\pm$ 0.032 & $-$ 6.345 $\pm$ 0.026 & $-$ 6.441 & $+$ 0.356 & $+$ 0.113 & 1.91 $\pm$ 0.14 &22\\
Cu I &  $-$ 7.670 $\pm$ 0.058 & $-$ 8.133 $\pm$ 0.067 & $-$ 7.941 & $+$ 0.271 & $-$ 0.028 & 2.17 $\pm$ 0.44 &3\\
Fe I &  $-$ 4.158 $\pm$ 0.038 & $-$ 4.439 $\pm$ 0.023 & $-$ 4.401 & $+$ 0.243 & $+$ 0.000 & 2.06 $\pm$ 0.11 &27\\
Mg I &  $-$ 4.228 $\pm$ 0.058 & $-$ 4.367 $\pm$ 0.095 & $-$ 4.441 & $+$ 0.213 & $-$ 0.030 & 1.50 $\pm$ 0.50 &3\\
Mn I &  $-$ 5.957 $\pm$ 0.097 & $-$ 6.600 $\pm$ 0.046 & $-$ 6.641 & $+$ 0.684 & $+$ 0.441 & 2.69 $\pm$ 0.21 &8\\
Na I &  $-$ 5.387 $\pm$ 0.048 & $-$ 5.789 $\pm$ 0.054 & $-$ 5.721 & $+$ 0.334 & $+$ 0.091 & 1.94 $\pm$ 0.31 &9\\
Ni I &  $-$ 5.367 $\pm$ 0.033 & $-$ 5.756 $\pm$ 0.027 & $-$ 5.821 & $+$ 0.454 & $+$ 0.211 & 2.29 $\pm$ 0.15 &17\\
Si I &  $-$ 4.111 $\pm$ 0.054 & $-$ 4.469 $\pm$ 0.058 & $-$ 4.401 & $+$ 0.290 & $+$ 0.047 & 2.39 $\pm$ 0.13 &23\\
Ti I &  $-$ 6.897 $\pm$ 0.040 & $-$ 7.064 $\pm$ 0.028 & $-$ 6.981 & $+$ 0.084 & $-$ 0.159 & 1.88 $\pm$ 0.13 &24\\
Zn I &  $-$ 7.028 $\pm$ 0.065 & $-$ 7.375 $\pm$ 0.048 & $-$ 7.441 & $+$ 0.413 & $+$ 0.170 & 2.25 $\pm$ 0.25 &4\\

 \hline 
 \end{tabular}
\end{table}

\begin{figure}[!h]
\centering
\begin{minipage}[t]{.75\linewidth}
\centering
\epsfig{file = 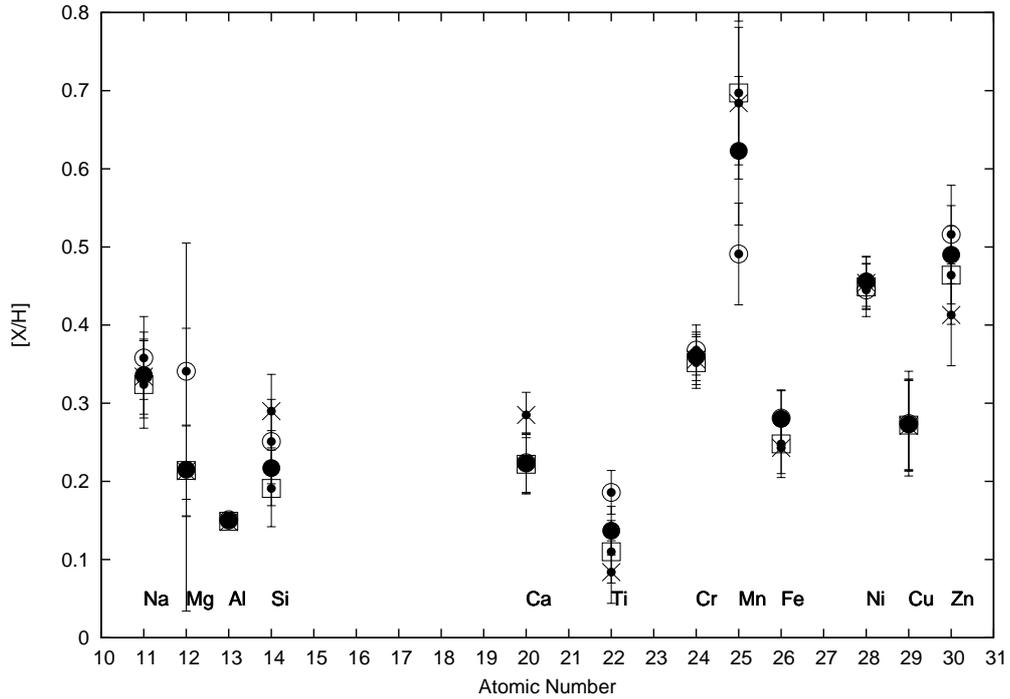,width = .7\linewidth, angle=270}
\caption{Dependence of [X/H] on atomic number of each element for HD 77338. Open circles show values found in the first iteration, filled circles are for
iteration 2, open squares are for iteration 3, and finally the stars show the results for iteration 4.}\label{fig4}
\end{minipage}
\end{figure}

\begin{figure}[!h]
\centering
\begin{minipage}[t]{.75\linewidth}
\centering
\epsfig{file = 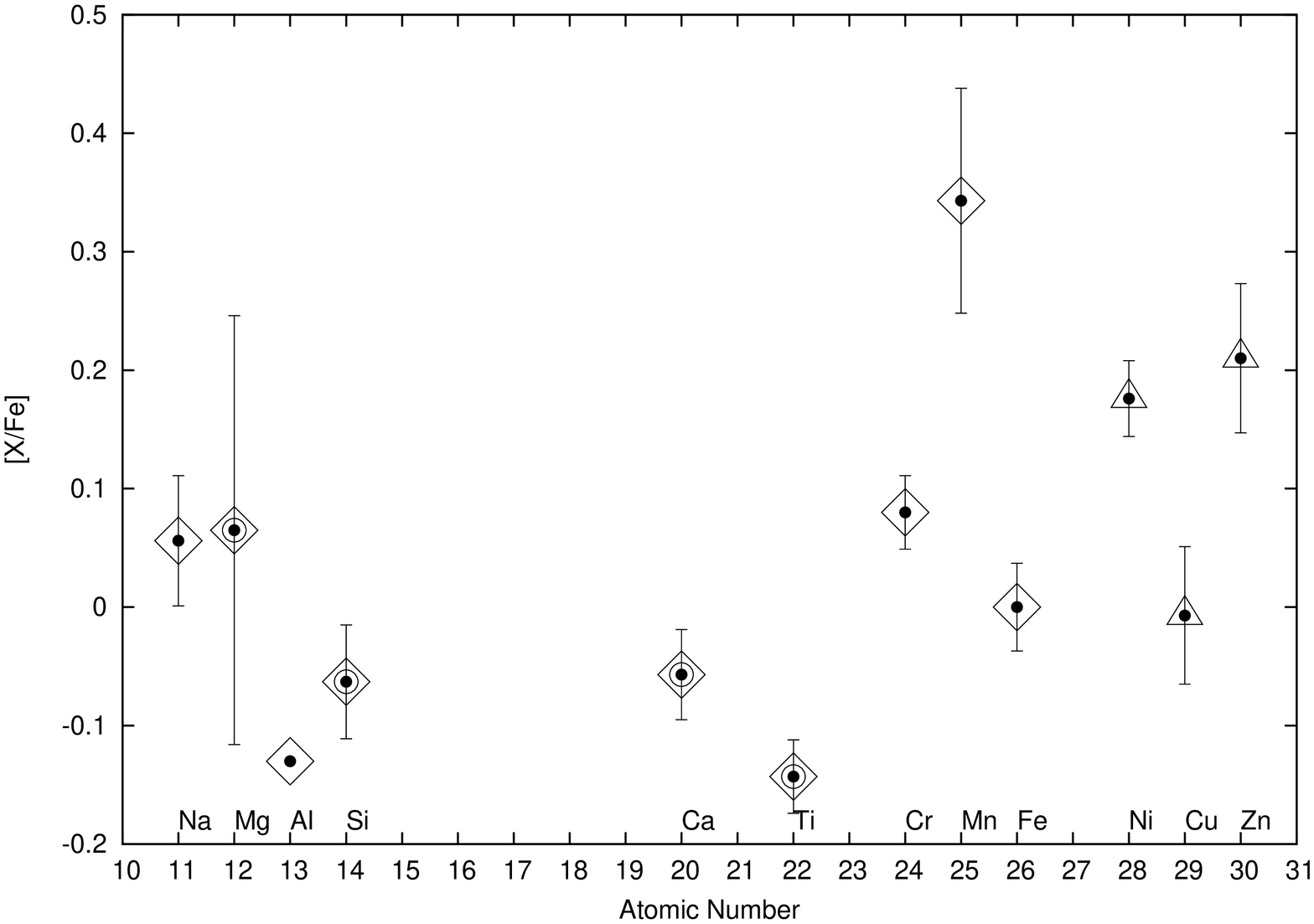,width = .7\linewidth,angle=270}
\caption{Dependence of [X/Fe] on the atomic number of each element. The different plotting shapes represent the different 
formation mechanism of each element. Open circles are for $\alpha$-elements, diamonds show the thermonuclear elemental production, and 
s-process is shown by triangles.}\label{fig5}
\end{minipage}
\end{figure}

  In Table~\ref{tab2} we present our results for 12 different ionic species.
We compare our abundances with the solar values, obtained using a 
model atmosphere of 5777/4.44/0.00. They are in good agreement with each other. 
Fig.~\ref{fig2} and Fig.~\ref{fig3} show the line profiles of Cr and Mg
calculated using a $V_{t}$ = 0.75 km/s.

In Fig.~\ref{fig4} we show the dependence of [X/H] on atomic number 
of each element for every iteration. The presence of errors can be explained by the presence of 
noise in the selected spectral lines, along with only having a small number of lines to work with for some elements. 
In Fig.~\ref{fig5} we present the dependence of [X/Fe] 
for the final iteration,  where different elements are shown using different
plotting shapes depending on their mechanism of formation.

\section{Discussion}

 We determined abundances for 12 ionic species in the atmosphere of the Sun and 
the metal-rich exoplanet host star HD 77338. Our values for the solar abundances are in good agreement with results 
from previous authors, proving the validity of our method. 
We used the solar spectrum as a reference to select the proper list of absorption lines to be 
used later in the analysis of the \HD spectrum.

Our [X/H] correlates well with the condensation temperature of the 
ions ($T_{cond}$), see discussion in \cite{_melendez}.
This may indicate the presence of a common shell (in the past) and can be an
additional criterion for the existence of a planetary system around metal-rich 
stars.

We also computed $ v \sin i $ for both stars. It is worth noting that
we determined all parameters in the framework of a fully self-consistent approach
(see \cite{SAM12} for more details).
In general, 
lines of Mn, Cu can not be used to obtain $ v \sin i $ because these lines
usually have several close components, but in our case the parameter 
$v \sin i$ was used to adjust theoretical profiles to get the proper 
fits to the observed special features. We believe that fits to Fe I lines
provide reasonable measures of the rotational velocity.

Our results show that the abundances of most elements in the atmosphere can be 
described well by the overall metallicity. However, we found an overabundance of 
some of the iron peak elements (e.g. Mn, Cu). Interestingly,
Cu is an element formed through the s-process and its abundance follows that of Fe, whereas Zn and 
elements formed through the p-process, e.g. Ni (see http://www.mao.kiev.ua/staff/yp/TXT/prs.png), 
show a noticeable overabundance compared to iron. It would be interesting to compare these results
for \HD with other metal rich stars to see if this is a common trend for super metal-rich stars. 
We plan to investigate this issue in a following paper (Ivanyuk et al. 2013, in preparation).

\section*{\sc acknowledgement}
\indent \indent JSJ acknowledges the support of the Basal-CATA grant.
YP's work has been supported
by an FP7 POSTAGBinGALAXIES grant
(No. 269193; International Research Staff Exchange
Scheme). 
Authors thank the
compilers of the international databases used
in our study:  SIMBAD (France, Strasbourg),
VALD (Austria, Vienna), and the authors of the atlas of the spectrum of the 
Sun as a star.
We thank anonymous Referee for some reasonable remarks and helpful comments.

\end{document}